\documentclass[preprint,amsmath,amssymb,nofootinbib]{revtex4}
\usepackage[dvips]{graphicx}
\usepackage{rotating}
\usepackage{hyperref}
\usepackage{subfig}
\usepackage{float}
\usepackage{array}
\usepackage[utf8]{inputenc}
\usepackage[normalem]{ulem}
\usepackage{color}
\definecolor{orange}{rgb}{1,0.5,0}
\usepackage[T1]{fontenc}

\newcommand{\tvect}[2]{\ensuremath{\Bigl(\negthinspace\begin{smallmatrix}#1\\#2\end{smallmatrix}\Bigr)}}

\begin{document}

\title{Angular and polarization trails from effective interactions of Majorana neutrinos at the LHeC}

\author{Luc\'{\i}a Duarte}
\email{lduarte@fing.edu.uy}
 \affiliation{Instituto de F\'{\i}sica, Facultad de Ingenier\'{\i}a,
 Universidad de la Rep\'ublica \\ Julio Herrera y Reissig 565,(11300) 
Montevideo, Uruguay.}

\author{Gabriel Zapata}
\author{Oscar A. Sampayo}
\email{sampayo@mdp.edu.ar}

 \affiliation{Instituto de Investigaciones F\'{\i}sicas de Mar del Plata 
(IFIMAR)\\ CONICET, UNMDP\\ Departamento de F\'{\i}sica,
Universidad Nacional de Mar del Plata \\
Funes 3350, (7600) Mar del Plata, Argentina.}

\begin{abstract}
We study the possibility of the LHeC facility to disentangle different new physics contributions to the production of heavy sterile Majorana neutrinos in the lepton number violating channel $e^{-}p\rightarrow l_{j}^{+} +  3 jets$ ($l_j\equiv e ,\mu $). This is done investigating the angular and polarization trails of effective operators with distinct Dirac-Lorentz structure contributing to the Majorana neutrino production, which parameterize new physics from a higher energy scale. We study an asymmetry in the angular distribution of the final anti-lepton and the initial electron polarization effect on the number of signal events produced by the vectorial and scalar effective interactions, finding both analyses could well separate their contributions. 
\end{abstract}

\maketitle

\section{\bf Introduction}

The discovery of neutrino masses through oscillation experiments continues to be the most compelling evidence of physics beyond the Standard Model (SM).
And yet the seesaw mechanism for neutrino mass generation \cite{Minkowski:1977sc, Mohapatra:1979ia, Yanagida:1980xy, GellMann:1980vs, Schechter:1980gr, Kayser:1989iu} plays a role as the most straightforward means to explain the tiny neutrino mass values. 
This mechanism introduces right handed sterile neutrinos $N_i$ which can have a Majorana mass term leading to the tiny known masses for the standard neutrinos, as long as the Yukawa couplings between the right handed Majorana neutrinos and the standard ones remain small. In fact, for Yukawa couplings of order $Y \sim 1$, a Majorana mass scale of order $M_{N} \sim 10^{15} GeV$ is needed to account for a light neutrino mass compatible with the current neutrino data ($m_{\nu}\sim 0.01 e$V)\cite{Patrignani:2016xqp}. On the other hand, for smaller Yukawa couplings, of order $Y\sim 10^{-8}-10^{-6}$, sterile neutrinos with masses around $M_{N}\sim (1-1000) ~GeV$ could exist, but in the simplest Type-I seesaw model with sterile Majorana neutrinos, this leads to negligible neutrino mixing values $U_{lN}^2 \sim m_{\nu}/M_N \sim 10^{-14}-10^{-10}$ \cite{Cai:2017mow, Atre:2009rg}. Thus, both alternatives lead to the decoupling of the Majorana neutrinos. In this scenario, the observation of lepton number violating (LNV) processes allowed by the existence of a Majorana neutrino mass term would be a sign of physics beyond the minimal seesaw mechanism \cite{delAguila:2008ir}.  

From the theoretical point of view, an alternative approach is to consider the Majorana neutrino interactions as originating in physics from a higher energy scale, parameterized by a model independent effective Lagrangian \cite{delAguila:2008ir, Duarte:2017rpc}. We consider that the sterile $N$ interacts with the SM particles by higher dimension effective operators, and take these interactions to be dominant in comparison with the mixing with light neutrinos through the Yukawa couplings, which we neglect. In this sense we depart from the usual viewpoint in which the sterile neutrinos mixing with the standard neutrinos is assumed to govern the $N$ production and decay mechanisms \cite{Atre:2009rg, delAguila:2007qnc}. 

The possibility for lepton number violation evidencing the Majorana nature of neutrinos in past and future hadron and lepton colliders has been extensively studied in the context of seesaw models (see \cite{Cai:2017mow, Deppisch:2015qwa} and references therein). The two-unit LNV channel $e^{-}p \rightarrow l_j^{+} +  3 jets$ has been studied in electron-proton colliders in \cite{Antusch:2016ejd, Blaksley:2011ey, Liang:2010gm, Ingelman:1993ve, Buchmuller:1991tu}. In this context, the LHeC proposed collider \cite{Bruening:2013bga, AbelleiraFernandez:2012cc} offers an opportunity to test the sterile neutrino interactions in this channel in a clean environment, as well as other interesting new physics models \cite{Curtin:2017bxr, Mondal:2015zba, Mondal:2016kof, Lindner:2016lxq}. 

The effective interactions we consider here for the heavy Majorana neutrinos were early studied in \cite{delAguila:2008ir}, where the possible phenomenology of dimension 6 effective operators was introduced. The dimension 5 operators extending the low-scale Type-I seesaw were investigated in \cite{Aparici:2009fh}. Their phenomenology is addressed in recent works as \cite{Ballett:2016opr, Caputo:2017pit}. Dimension 7 effective $N$ operators are studied in \cite{Bhattacharya:2015vja, Liao:2016qyd}. The collider phenomenology of the dimension 6 effective Lagrangian used in this paper has been studied by our group and others in \cite{delAguila:2008ir, Peressutti:2011kx, Peressutti:2014lka, Duarte:2014zea, Duarte:2015iba, Duarte:2016miz, Duarte:2016caz, Yue:2017mmi}.

The different operators in the effective Lagrangian parameterize a wide variety of UV-complete new physics models, like extended scalar and gauge sectors as the Left-Right symmetric model, vector and scalar leptoquarks, etc. Thus, discerning between the possible contributions given by them to specific processes gives us a hint on what kind of new physics at a higher energy regime is responsible for the observed interactions. 

In \cite{Duarte:2014zea} we studied the potential of the LHeC to discover Majorana neutrinos for different values of their mass, effective couplings and the new physics scale. Here we aim to go further in the study of the Majorana neutrino effective interactions, and point towards disentangling the possible contributions of effective operators with different Dirac-Lorentz structure to the $e^- p\rightarrow l^{+} +  3 jets$ process with the aid of angular distributions and polarization effects. 

We study the influence of vectorial and scalar operators on the angular distribution of the final anti-lepton, building a forward-backward asymmetry, and study the potential of using the initial electron polarization as a discriminator between both effective operator groups. 

The paper is organized as follows. In Sec. \ref{sec:modelo} we review the effective Lagrangian and the existing constraints on the values of the effective couplings. In Sec. \ref{sec:signal} we show the analytic and numerical signal calculation, as well as the cuts imposed for background suppression. The numerical results for the final anti-lepton angular distributions and forward-backward asymmetry are presented in Sec. \ref{sec:angular}, and the initial electron polarization analysis is presented in Sec. \ref{sec:pol}. Our conclusions are derived in Sec. \ref{sec:Concl}.

\section{\bf Majorana neutrino interaction model \label{sec:modelo}}

\subsection{Effective operators and Lagrangian}

As the heavy sterile Majorana neutrino $N$ is a SM singlet, its only possible renormalizable interactions with SM fields involve the Yukawa couplings, which must be very small in order to accommodate the observed tiny ordinary $\nu$ masses. Thus, any observation of leptonic number non-conservation should be a manifestation of physics beyond the minimal see-saw mechanism \cite{delAguila:2008ir}. Our aim is to investigate the possible contributions of a heavy Majorana neutrino with negligible mixing to the SM $\nu_{L}$. We consider the most simple scenario, including only one heavy neutrino state $N$ as an observable degree of freedom. 

The effects of the new physics involving one heavy sterile neutrino and the SM fields are parameterized by a set of effective operators $\mathcal{O}_\mathcal{J}$ satisfying the $SU(2)_L \otimes U(1)_Y$ gauge symmetry \cite{Wudka:1999ax}. 
The contribution of these operators to observable quantities is suppressed by inverse powers of the new physics scale $\Lambda$. The total Lagrangian is organized as follows:

\begin{eqnarray}\label{eq:Lagrangian}
\mathcal{L}=\mathcal{L}_{SM}+\sum_{n=5}^{\infty}\frac1{\Lambda^{n-4}}\sum_{\mathcal{J}} \alpha_{\mathcal{J}} \mathcal{O}_{\mathcal{J}}^{(n)}
\end{eqnarray}
where $n$ is the mass dimension of the operator $\mathcal{O}_{\mathcal{J}}^{(n)}$.

Note that we do not include the Type-I seesaw Lagrangian -the Majorana and Yukawa terms- giving rise to the mixing between the sterile and the standard left-handed neutrinos, which we are neglecting. In this work it is considered that the dominating new physics effects leading to the lepton number violation come from the lower dimension operators that can be generated at tree level in the unknown underlying renormalizable theory.

The dimension 5 operators in \eqref{eq:Lagrangian} were studied in detail in \cite{Aparici:2009fh}. These include the well known Weinberg operator $\mathcal{O}_{W}\sim (\bar{L}\tilde{\phi})(\phi^{\dagger}L^{c})$ \cite{Weinberg:1979sa} contributing to the light neutrino masses, and operators with the $N$: $\mathcal{O}_{N\phi}\sim (\bar{N}N^{c})(\phi^{\dagger} \phi)$ contributing to the $N$ Majorana masses and giving couplings of the heavy neutrinos to the Higgs (its phenomenology for the LHC has been studied very recently in \cite{Caputo:2017pit}), and an operator $\mathcal{O}^{(5)}_{NB}\sim (\bar{N}\sigma_{\mu \nu}N^{c}) B^{\mu \nu}$ inducing magnetic moments for the heavy neutrinos, which is identically zero if we include just one sterile neutrino $N$ in the theory\footnote{The effects of considering the $\mathcal{O}^{(5)}_{NB}$ operator were studied in \cite{Aparici:2009fh} for the case of 2 massive Majorana neutrinos $N_{1,2}$. Our treatment coincides with the limit in which $N_{1,2}$ are mass-degenerate and the light-heavy mixing is taken to be zero.}. In the following, as the dimension 5 operators do not contribute to the studied processes -discarding the heavy-light neutrino mixings- we will only consider the contributions of the dimension 6 operators, following the treatment made in \cite{delAguila:2008ir}.

We organize the effective operators in different subsets.
The first one includes operators with scalar and vector bosons (SVB),
\begin{eqnarray} \label{eq:ope1}
\mathcal{O}_{LN\phi}=(\phi^{\dag}\phi)(\bar L N \tilde{\phi}), \;\; \mathcal{O}_{NN\phi}=i(\phi^{\dag}D_{\mu}\phi)(\bar N
\gamma^{\mu} N), \;\; \mathcal{O}_{Ne\phi}=i(\phi^T \epsilon D_{\mu} \phi)(\bar N \gamma^{\mu} l)
\end{eqnarray}
and a second subset includes the baryon-number conserving 4-fermion contact terms:
\begin{eqnarray} \label{eq:ope2}
\mathcal{O}_{duNe}&=&(\bar d \gamma^{\mu} u)(\bar N \gamma_{\mu} l) , \;\; \mathcal{O}_{fNN}=(\bar f \gamma^{\mu}
f)(\bar N \gamma_{\mu}
N), \;\; \mathcal{O}_{LNLe}=(\bar L N)\epsilon (\bar L l),
\nonumber \\
\mathcal{O}_{LNQd}&=&(\bar L N) \epsilon (\bar Q
d), \;\; \mathcal{O}_{QuNL}=(\bar Q u)(\bar N L) , \;\; \mathcal{O}_{QNLd}=(\bar Q N)\epsilon (\bar L d) 
\end{eqnarray}
where $e_i$, $u_i$, $d_i$ and $L_i$, $Q_i$ denote, for the family
labeled $i$, the right handed $SU(2)$ singlet and the left-handed
$SU(2)$ doublets, respectively. Here $ \gamma^{\mu}$ are the Dirac matrices, and $\epsilon=i\sigma^{2}$ is the antisymmetric symbol. We do not consider the one-loop generated operators which are naturally suppressed by a factor $1/16\pi^2$ \cite{delAguila:2008ir, Arzt:1994gp}. The complete expression for the dimension 6 effective Lagrangian can be found in an appendix in \cite{Duarte:2016miz}.

The effective operators in \eqref{eq:ope1} and \eqref{eq:ope2} cover a wide variety of new physics models, as we mentioned in the introduction. The effects of the four-fermion contact operators $\mathcal{O}_{duNe}\equiv \mathcal{O}_{V_{0}}$ and $\mathcal{O}_{QNLd}\equiv \mathcal{O}_{S_{3}}$ have been studied recently as a parameterization of the minimal Left-Right Symmetric Model (LRSM) in recasts of LHC searches for the same-sign dilepton signal \cite{Ruiz:2017nip}.

\begin{figure*}[h]
 \includegraphics[totalheight=5.8cm]{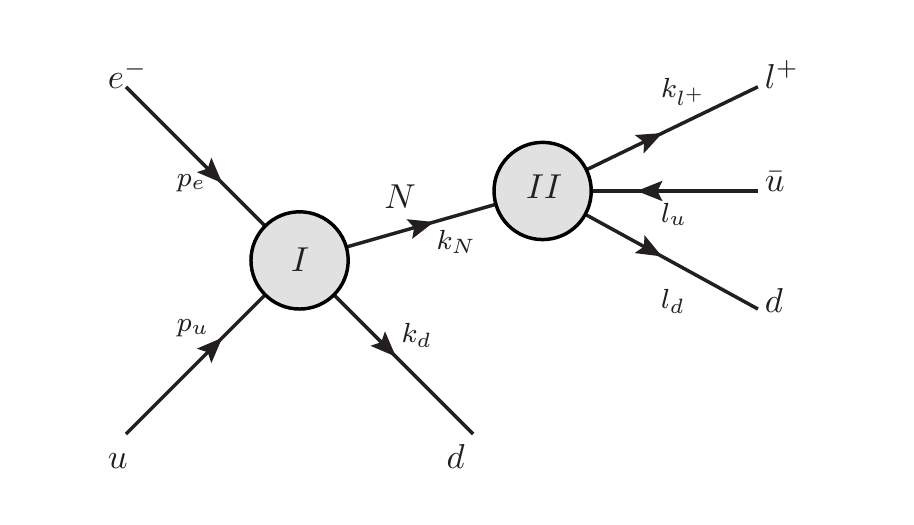}
\caption{\label{fig:epl} Process $e p \rightarrow l^{+} + 3jets + X$ with N decaying according to Ref.\cite{Duarte:2016miz, Duarte:2015iba} }
\end{figure*}

In order to obtain the interactions involved in the process $ep \rightarrow l^{+} +3 jets$ depicted in Fig.\ref{fig:epl} we consider the effective Lagrangian terms involved in the calculations, taking the scalar doublet after spontaneous symmetry breaking as $\phi=\tvect{0}{\frac{v+h}{\sqrt{2}}}$, with $h$ being the Higgs field and $v$ its v.e.v. For the Majorana neutrinos production $(I)$ and decay $(II)$ vertices in Fig.\ref{fig:epl} we have contributions to the effective Lagrangian related to the spontaneous symmetry breaking process coming from \eqref{eq:ope1} and the four-fermion interactions involving quarks and leptons from \eqref{eq:ope2}:
\begin{eqnarray}\label{eq:leff}
\mathcal{L}_{eff}&=&\frac{1}{\Lambda^2}\left\{- \frac{m_W v}{\sqrt{2}} 
\alpha^{(i)}_W
\; W^{\dag\; \mu} \; \overline N_R \gamma_{\mu} e_{R,i} + \alpha^{(i,j)}_{V_0} 
\bar d_{R,i} \gamma^{\mu} u_{R,i} \overline N_R \gamma_{\mu}
e_{R,j} + \right.
\nonumber
\\ &&
 \alpha^{(i,j)}_{S_1}(\bar
u_{L,i}u_{R,i}\overline N \nu_{L,j}+\bar d_{L,i}u_{R,i} \overline N e_{L,j})
 +
\alpha^{(i,j)}_{S_2} (\bar \nu_{L,i}N_R \bar d_{L,i}d_{R,j}-\bar
e_{L,i}N_R \bar u_{L,i}d_{R,j}) +
\nonumber
\\ &&
\left. \alpha^{(i,j)}_{S_3}(\bar u_{L,i}N_R
\bar e_{L,i}d_{R,j}-\bar d_{L,i}N_R \bar \nu_{L,i}d_{R,j})
  + h.c. \right\}
\end{eqnarray}
where the sum over the fermion families $i,j=1,2,3$ is understood and the couplings
$\alpha^{(i,j)}_{\mathcal{J}}$ are associated to specific operators according to
\begin{eqnarray}\label{eq:alphas}
\alpha_W=\alpha_{Ne\phi},\;
\alpha_{V_0}=\alpha_{duNe},\;\;
\alpha_{S_1}=\alpha_{QuNL},\;
\alpha_{S_2}=\alpha_{LNQd},\;\;
\alpha_{S_3}=\alpha_{QNLd}~.\;
\end{eqnarray}
In this work we allow for family mixing in the interactions involving two or more different SM leptons: this allows for the appearance of $\mu^{+}$ anti-leptons together with positrons in the final state. The case $l^+=\tau^+$ is allowed in theory, but we do not take it into account, due to the difficult tau reconstruction in experiments.

The effective operators above can be classified by their Dirac-Lorentz structure into {\it{scalar}} and {\it{vectorial}}. The scalar and vectorial operators contributing to the studied processes are those appearing in \eqref{eq:leff} with couplings named $\alpha_{S_{1,~2,~3}}$ and $\alpha_{W,~ V_{0}}$, respectively.

The relative sizes between the different effective couplings are given by the contribution of the corresponding operators to the experimental observables.

\subsection{Effective coupling bounds summary}\label{subsec:bounds}

The current experimental constraints on the light-heavy neutrino mixing parameters in seesaw models can be exploited to impose plausible constraints on the numerical values of the effective couplings $\alpha_{\mathcal{J}}$, which weight the relative importance of the possible effective interactions.

In the literature \cite{deGouvea:2015euy, Deppisch:2015qwa, Antusch:2015mia, Das:2014jxa, Fernandez-Martinez:2016lgt} the existing experimental bounds are summarized in general phenomenological approaches considering low scale minimal seesaw models, parameterized by a single heavy neutrino mass scale $M_{N}$ and a light-heavy mixing $U_{lN}$, with $l$ indicating the lepton flavor. We exploit these results -like constraints on processes as neutrino-less double beta decay ($0\nu\beta\beta$), electroweak precision data (EWPD), LNV rare meson decays as well as direct collider searches, including $Z$ decays- linking the $U_{lN}$ mixings in Type-I seesaw models \cite{Atre:2009rg, delAguila:2007qnc} with our effective couplings in \ref{eq:Lagrangian} by the ad-hoc relation
\begin{equation}
U^2_{lN}\simeq \left(\frac{\alpha v^2}{2\Lambda^2}\right)^2.
\end{equation}
In previous works \cite{Duarte:2016miz, Duarte:2015iba} we have presented our procedure, and refer the reader to those papers for a detailed discussion. 

For the couplings involving the first fermion family -taking indices $i=1$ and $j=1$ in \eqref{eq:leff}- the most stringent are the $0\nu\beta\beta$-decay bounds obtained by the KamLAND-Zen collaboration \cite{KamLAND-Zen:2016pfg}. Following the treatment made in \cite{Mohapatra:1998ye, deGouvea:2015euy, Duarte:2016miz}, they give us an upper limit $\alpha^{bound}_{0\nu\beta\beta} \leq 3.2 \times 10^{-2} \left(\frac{m_N}{100 ~GeV}\right)^{1/2}$, where the new physics scale is taken to be $\Lambda=1~TeV$ (here and in the following) \footnote{The new physics scale $\Lambda=1~TeV$ is taken as an illustration. One can obtain the values at any other scale $\Lambda^{\prime}$ considering $\alpha^{\prime}_{\mathcal{J}}=(\frac{\Lambda^{\prime}}{\Lambda})^{2} \alpha_{\mathcal{J}}$. }. For the second fermion family -taking indices $i=2$ or $j=2$ in \eqref{eq:leff}- and sterile neutrino masses in the range $m_W\lesssim m_N$ the upper limits come from EWPD like radiative lepton flavor violating (LFV) decays as $\mu\rightarrow e \gamma$ \cite{Tommasini:1995ii, delAguila:2008pw, deGouvea:2015euy, Fernandez-Martinez:2016lgt} giving a bound $\alpha^{bound}_{EWPD} \leq 0.32$. 

In our numerical analysis throughout the paper -for the sake of simplicity- we take the couplings associated to the operators that contribute to the $0\nu\beta\beta$-decay for the first family ($i,j=1$) as restricted by the corresponding bound $\alpha^{bound}_{0\nu\beta\beta}$, and we fix the other couplings to the value $\alpha^{bound }\leq 0.32$ valid for high Majorana neutrino masses\footnote{Allowing for family mixing does not impose severe bounds in the high $m_N$ range we are considering in this paper.}, as detailed in Tab.\ref{tab:alpha-sets}.  

\begin{table}[t]
 \centering
 \begin{tabular}{c c c}
\firsthline
 $\alpha^{bound}_{0\nu\beta\beta}= 3.2 \times 10^{-2} 
\left(\frac{m_N}{100 ~GeV}\right)^{1/2}$, ~$\Lambda=1 ~TeV$,  & & $\alpha^{bound}_{EWPD}= 0.32$\\
\hline
$\alpha^{(1)}_{W}=\alpha^{bound}_{0\nu\beta\beta}$& & $\alpha^{(2)}_{W}= \alpha^{bound}_{EWPD}$  \\
$\alpha^{(1,1)}_{V_{0}}=\alpha^{bound}_{0\nu\beta\beta}$ & & $\alpha^{(1,2)}_{V_{0}}=\alpha^{(2,1)}_{V_{0}}=\alpha^{(2,2)}_{V_{0}}=\alpha^{bound}_{EWPD}$ \\
$\alpha^{(1,1)}_{S_{1,2,3}}=\alpha^{bound}_{0\nu\beta\beta}$ & & $\alpha^{(1,2)}_{S_{1,2,3}}=\alpha^{(2,1)}_{S_{1,2,3}}=\alpha^{(2,2)}_{S_{1,2,3}}=\alpha^{bound}_{EWPD}$ \\
\lasthline
 \end{tabular}
\caption{Effective couplings numerical values. }\label{tab:alpha-sets}
\end{table}

\section{\bf Signal detection}\label{sec:signal}

The LHeC is proposed to be an $e^{-}p$ collider built at the LHC tunnel, using an electron beam in the $60-150 ~GeV$ energy range with the existing $7~TeV$ proton beam. It is expected to achieve an integrated luminosity $\mathcal{L}=100 ~fb^{-1}$ per operation year. One possibly very crucial feature of the LHeC is the availability of a polarized electron beam \cite{Bruening:2013bga, AbelleiraFernandez:2012cc}, which has already been studied as a chance to enhance the observability of Majorana neutrinos in the context of the Left-Right symmetric SM extension \cite{Mondal:2015zba, Lindner:2016lxq}.  

In this paper we study the effects of the possible existence of a heavy sterile Majorana neutrino $N$ with effective interactions on the angular distribution of the produced anti-lepton and the effects of the initial electron polarization in the process $ep \rightarrow l_{j}^{+} +  3 jets$ ($l_j\equiv e ,\mu $) at the LHeC.

The cross section for the process $e p \rightarrow l^{+} + 3 jets$ is calculated using the Lagrangian 
in \eqref{eq:leff}, according to the process in Fig. \ref{fig:epl}. The analytical expression is
\begin{eqnarray}
\sigma(ep\rightarrow l^+ + 3 jets)=\sum_i \int_{m_N^2/s}^1 dx f_i(x) \hat 
\sigma_i(x s)
\end{eqnarray}
where the center of mass energy is taken to be $\sqrt{s}=\sqrt{4 E_e E_p}$, $\hat{\sigma}$ 
is the parton level scattering cross section and $\hat{s}$ the squared center of mass energy. 
Here $i=1$ corresponds to the channel $e u \rightarrow N d$ and $i=2$ to the 
crossed channel $e \bar d \rightarrow N \bar u$. The function $f_1(x)$ represents the $u(x)$ parton distribution function (PDF), 
and $f_2(x)$ the one for $\bar d(x)$. For numerical calculations we use the CTEQ set \cite{Pumplin:2002vw}.

The parton level cross section is written as
\begin{eqnarray}
\label{eq:sigma}
\hat \sigma_i(x s)= \int (2\pi)^4 
\delta^{(4)}(p_e+p_u-\sum_{j=1,4} 
k_j)\overline{|M_{(i)}|^2} \prod_{j=1,4}\frac{d^4k_j}{2\pi^3} ~,
\end{eqnarray}
and the squared scattering amplitudes in the narrow width approximation are :
\begin{eqnarray}
\overline{\vert M_{(i)} \vert}^2=\left(\frac{\pi}{4m_N ~
\Gamma_N ~\hat{s}}\right)\delta(k_N^2-m_N^2) \vert \Lambda_{(I),i} \vert^2 
(\vert \Lambda_{(II)}^{(+)} \vert^2 + \vert \Lambda_{II}^{(-)} \vert^2)
\end{eqnarray}
where \footnote{Here and in Sec. \ref{sec:pol} we omit the family superscripts in the effective couplings for simplicity.}
\begin{eqnarray}\label{eq:sigLam}
\vert \Lambda_{(I),1} \vert^2 &&= \frac{4}{\Lambda^2}\left[ 
4(\alpha_{S_{2}}(\alpha_{S_{2}}-\alpha_{S_{3}})+\alpha^2_{S_{1}}) (k_d \cdot 
p_u)(k_N \cdot p_e)+          
\right. \nonumber \\  &&\left. (4 \alpha_{W}^2 \vert \Pi^{(2)}_W \vert^2 
+\alpha_{S_{3}}(\alpha_{S_{3}}-\alpha_{S_{2}}))(k_d \cdot p_e)(k_N \cdot p_u)+ 
(\alpha_{S_{3}} \alpha_{S_{2}}+4 \alpha^ 2_{V_{0}}) (k_d \cdot k_N)(p_e \cdot 
p_u) \right]
\nonumber\\
\nonumber\\
\vert \Lambda_{(II)}^{(-)} \vert^2 &&= \frac{16}{\Lambda^4}\left[\vert 
\Pi^{(2)}_W \vert^2 \alpha_W^2 
(k_N \cdot l_u)(k_{l^+} \cdot l_d)+ \alpha_{V_0}^2 (k_N \cdot l_d)(k_{l^+} 
\cdot l_u) \right]
\nonumber\\
\nonumber\\
\vert \Lambda_{(II)}^{(+)} \vert^2 &&= 
\frac{4}{\Lambda^4}\left[(\alpha_{S_1}^2+\alpha_{S_2}^2-\alpha_{S_2}\alpha_{S_3}
)(l_u \cdot l_d)(k_{l^+} \cdot k_N) + \right.
\nonumber \\ &&\left. (\alpha_{S_3}^2-\alpha_{S_2}\alpha_{S_3})(k_{l^+} \cdot  
l_d)(l_u \cdot k_N)+
\alpha_{S_2} \alpha_{S_3} (l_u \cdot k_{l^+})(l_d \cdot k_N) \right]
\end{eqnarray}
with $\Pi^{(1)}_W=m_W^2/(-2(p_u \cdot k_d)-m_W^2)$, $\Pi^{(2)}_W=m_W^2/(2(l_u \cdot l_d)-m_W^2)$. The final leptons can be either $e^{+}$ or $\mu^{+}$ since this is allowed by the interaction Lagrangian \eqref{eq:leff}. These final states are clear signals for intermediary Majorana neutrinos, thus we sum the cross section over the flavors of the final leptons. The total width ($\Gamma_N$) for the Majorana neutrino decay is calculated in \cite{Duarte:2016miz}.

The numerical cross section for the Majorana neutrino production in $e p$ colliders and the following decay $N \rightarrow l^+ + 2 jets$ is updated from \cite{Duarte:2014zea}, considering the values for the effective couplings in Tab. \ref{tab:alpha-sets}, and the full $N$ decay width calculated in \cite{Duarte:2016miz}.

\begin{figure*}[h]
\centering
\subfloat[Signal cross section. Effective numerical values shown in Tab.\ref{tab:alpha-sets}.]{\label{fig:sigma}\includegraphics[totalheight=5.8cm]{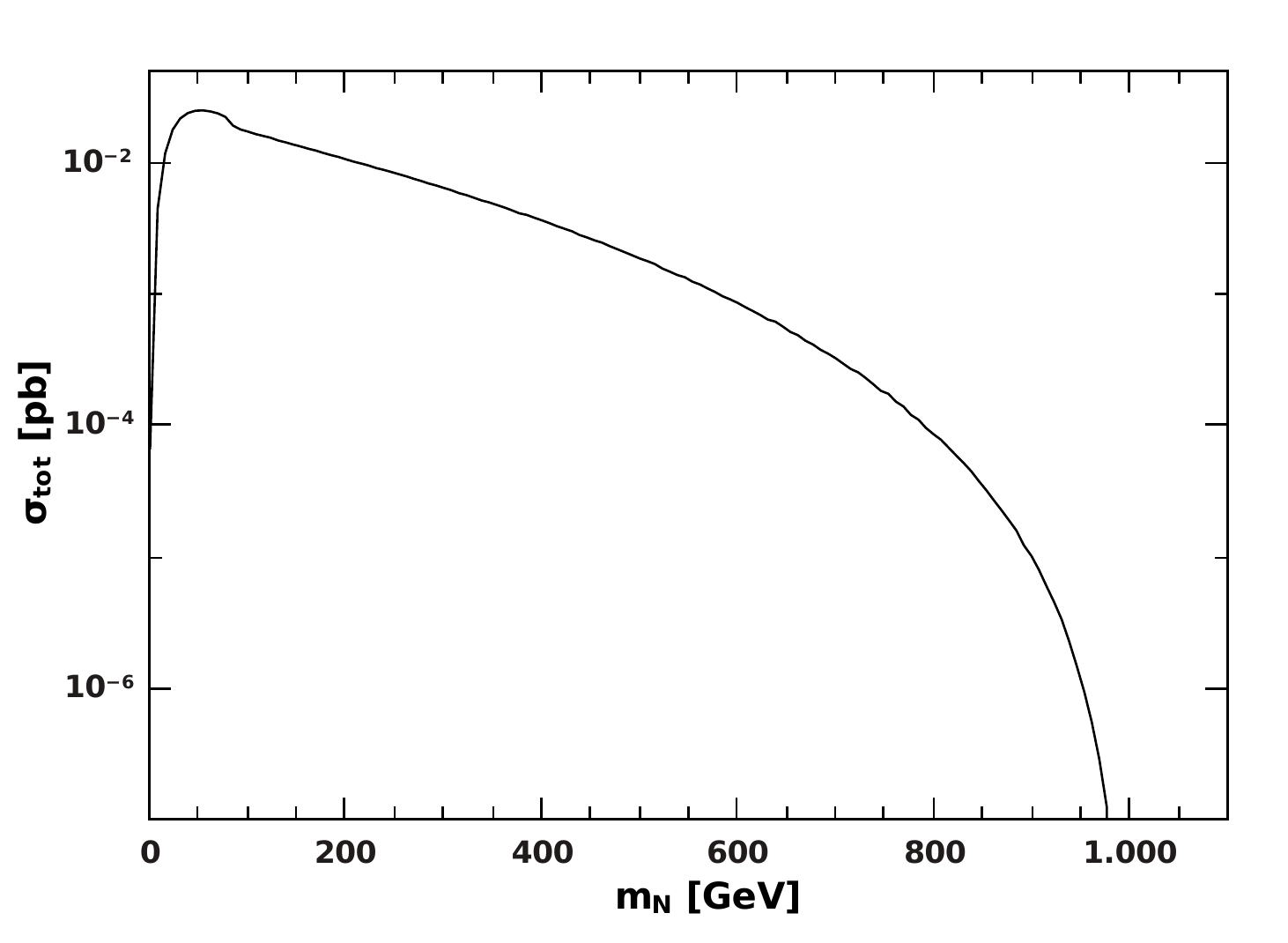}}~
\subfloat[Background $E_T$ dependence.]{\label{fig:cutEt}\includegraphics[totalheight=5.8cm]{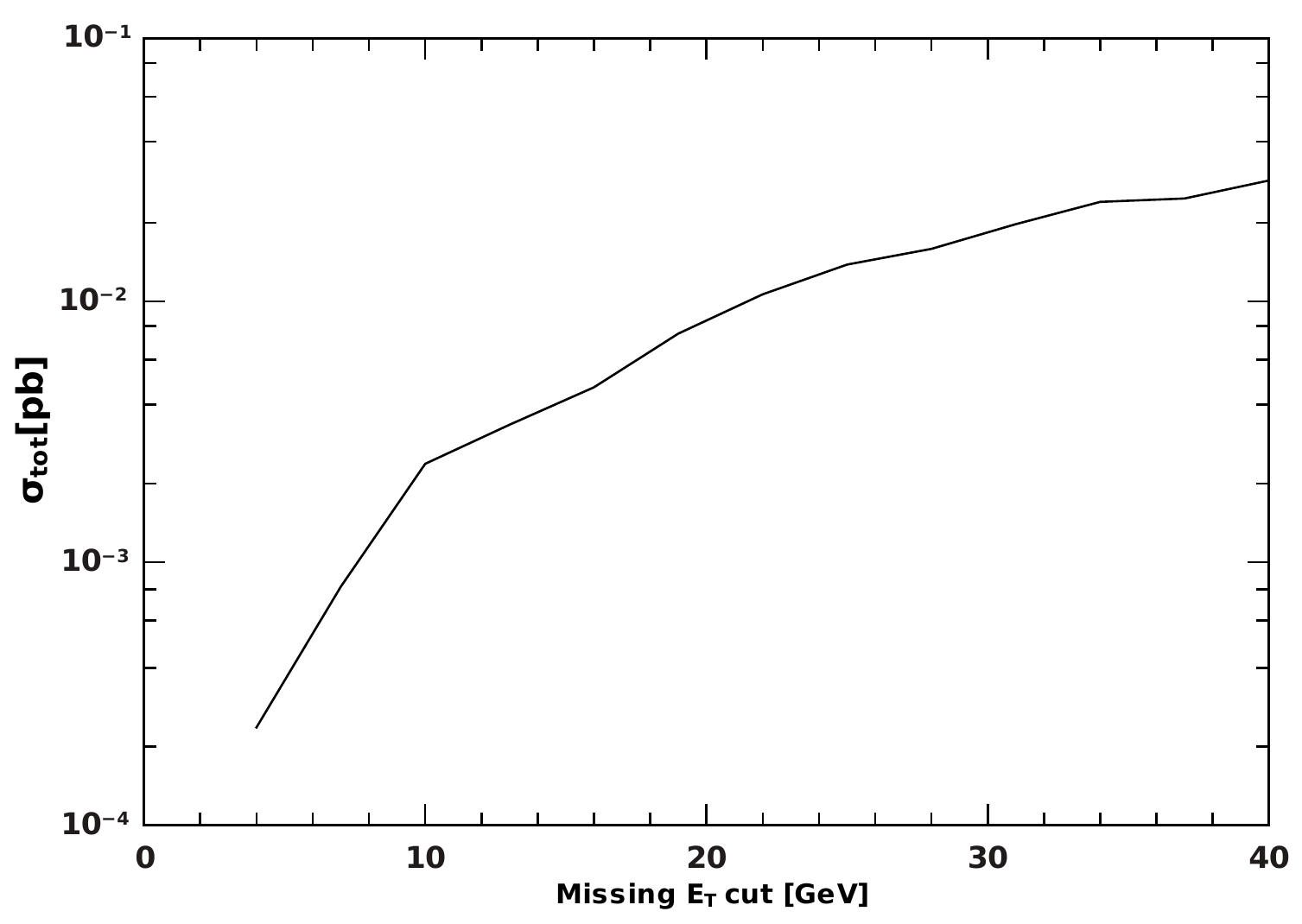}}
\caption{Cross section for the process $e p \rightarrow l^{+} + 3jets$ with N decaying according to Ref.\cite{Duarte:2016miz, Duarte:2015iba}.  (a) and background dependence with missing $E_T$ (b).}
\end{figure*}

In Fig.\ref{fig:sigma} we show the results for the cross section, as a function of the Majorana neutrino mass $m_N$, for the electron beam energy $E_{e}=150 ~GeV$ and for a $E_P=7 ~TeV$ proton beam. We have considered $\sqrt{\hat{s}} < \Lambda$ in order to ensure the validity of the effective Lagrangian approach. The effective couplings are taken as in Tab. \ref{tab:alpha-sets}. The phase space integration of the squared amplitude is made generating the final momenta with the Monte Carlo routine RAMBO \cite{Kleiss:1985gy}.

Although the lepton number violating considered signal is strictly forbidden in the SM, the backgrounds for the studied process were carefully investigated in \cite{Blaksley:2011ey, Duarte:2014zea}. The dominant background comes from $W \rightarrow l^+$ ($e^+,\mu^+$) events. The process $e^{-}p\rightarrow e^- l^+ j j j \nu$ is not distinguished from the signal if the outgoing electron is lost in the beamline, and is dominated by the exchange of an almost real photon with a very collinear outgoing electron ($p\gamma\rightarrow l^+  + 3j +\nu$), which convoluted with the PDF representing the probability of finding a photon inside an electron, is found to be the major contribution to $W$ production. In this paper we have updated the simulation of background processes, done with CalcHep \cite{Belyaev:2012qa}.

\begin{figure*}
\centering
\includegraphics[scale=0.9]{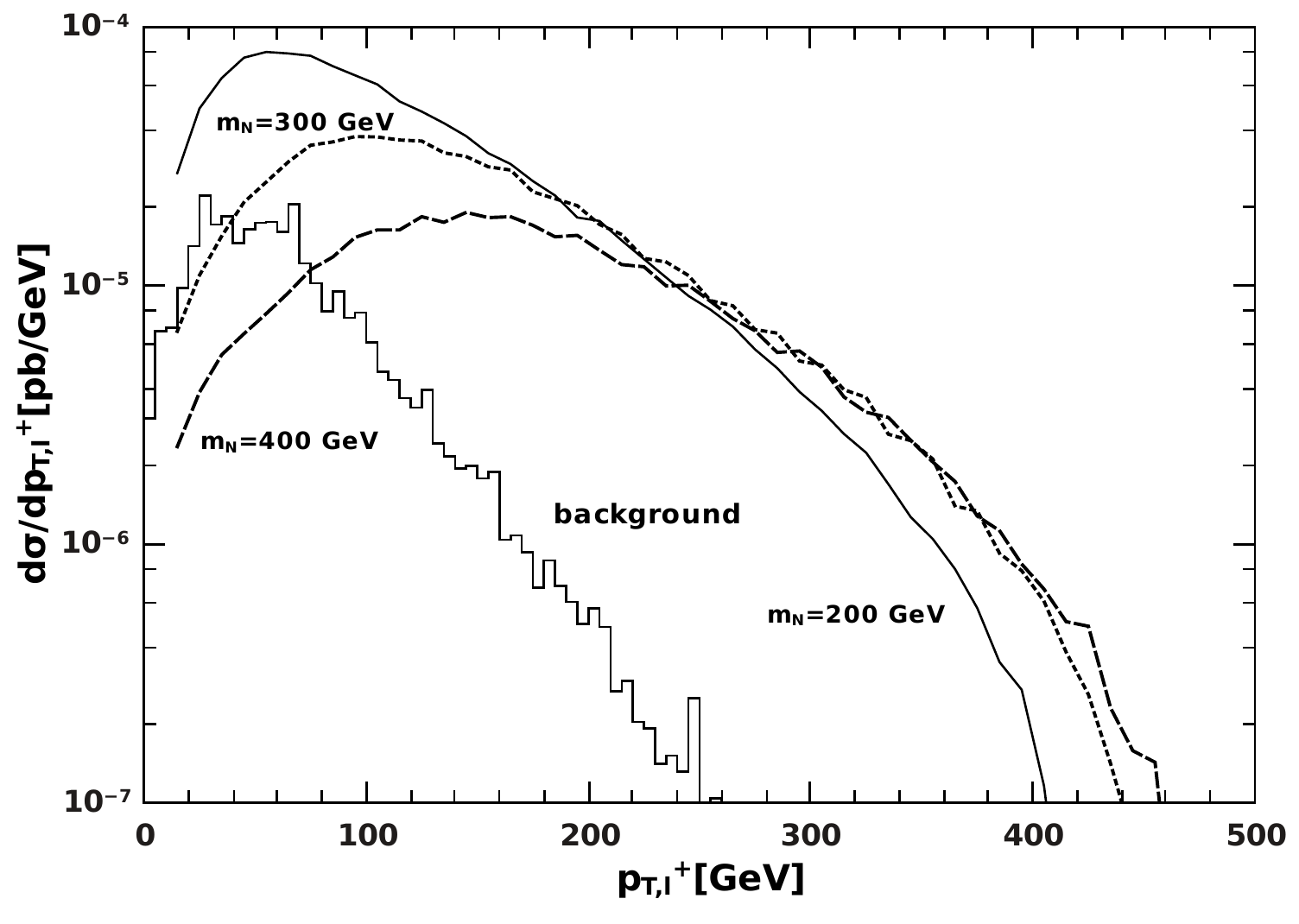}
\caption{\label{fig:dptcutet} Differential cross section of signal and 
background in function of transverse momentum $p_{T,l^+}$. The cut in missing $E_{T}$ 
is included.}
\end{figure*}

As the SM background always involves final state neutrinos, in \cite{Blaksley:2011ey, Duarte:2014zea} was found that a cut on the missing $E_{T}$ helps in reducing this background. Other efficient cut is on the $l^ +$ minimum transverse momentum. In Fig.\ref{fig:cutEt} we show the behavior of the background with the maximum missing energy $E_T$ for $E_e=150 ~GeV$. A cut of $E_{T,max}\le 10$ GeV, which is a reasonable value for the detector resolution, has not appreciable effects on the signal but reduces the background significantly. In Fig.\ref{fig:dptcutet} we show the differential cross section for the 
background and the signal for different values of the Majorana masses as a function of the transverse momentum $p_{T,l^+}$ of the anti-lepton. In this figures the cut on the missing energy $E_T$ has already been included. It can be seen that the background is mostly concentrated at low values of $p_{T,l^+}$, and a cut imposed on $p_{T,l^+}^{min}$ could be effective to improve the signal/background relation. 

\begin{figure*}
\includegraphics[scale=0.9]{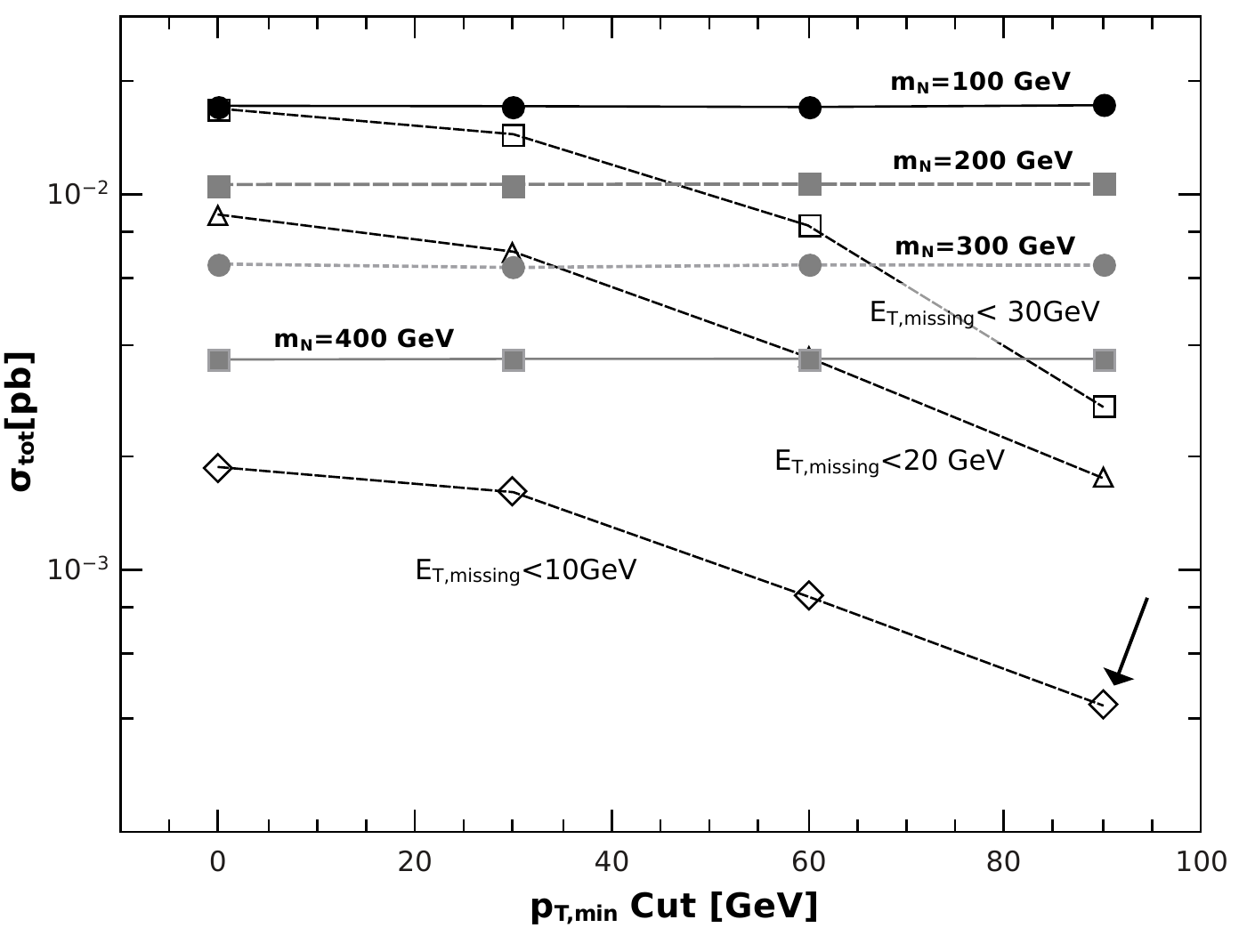}
\caption{\label{fig:sigbck} Comparison between signal and background for 
different Majorana neutrino masses, cut in missing $E_T$ and the transversal 
momentum of the final lepton $p_{T,l^ +}$. The arrows 
indicate the cuts and backgrounds used in the analysis.}
\end{figure*}

Finally, in Fig.\ref{fig:sigbck} we show a plot comparing the magnitude of the signal for different values of the Majorana neutrino mass, and the background for different $E_{T,max}$ cuts (black dashed lines), depending on the $p_{T,l^+}^{min}$ cut imposed. In the figure the arrows indicate the value of the cuts used in the analysis: we impose $p_{T,l^+} \ge 90$ GeV and $E_{T,miss} \le 10$ GeV in order to reduce the background without appreciably decreasing the signal.

\section{\bf Angular distributions and asymmetry}\label{sec:angular}

The use of angular asymmetries to disentangle the contributions from effective operators with different Lorentz-Dirac structure was already proposed in \cite{delAguila:2008ir}. A forward-backward like asymmetry was studied in \cite{Duarte:2016caz} for the case of long-lived Majorana neutrinos in the well known di-lepton LNV channel at the LHC. Recently, a forward-backward asymmetry is used to disentangle the Dirac or Majorana nature of intermediate neutrinos in purely leptonic $N$ decays at the LHC \cite{Arbelaez:2017zqq}.

With the final anti-lepton angular distribution $d\sigma/d\cos\theta$ where $\theta$ is the angle between the outgoing anti-lepton and the incident electron beam in the lab frame, we construct a forward-backward asymmetry 
$A^{l^+}_{FB}$ as a function of the number of events in each hemisphere:
\begin{eqnarray}
A^{l^+}_{FB}=\frac{N_+ \; - \; N_-}{N_+ \; + \; N_-}
\end{eqnarray}
where $N_+$ is the number of events with an angle in the interval $0\leq \theta \leq \pi/2$ and $N_-$ the number of events with $\pi/2\leq \theta \leq \pi$. 
\begin{figure*}
\centering
\subfloat[Vectorial operators contribution.]{\label{fig:dsdcos_vec}\includegraphics[totalheight=6.5cm]{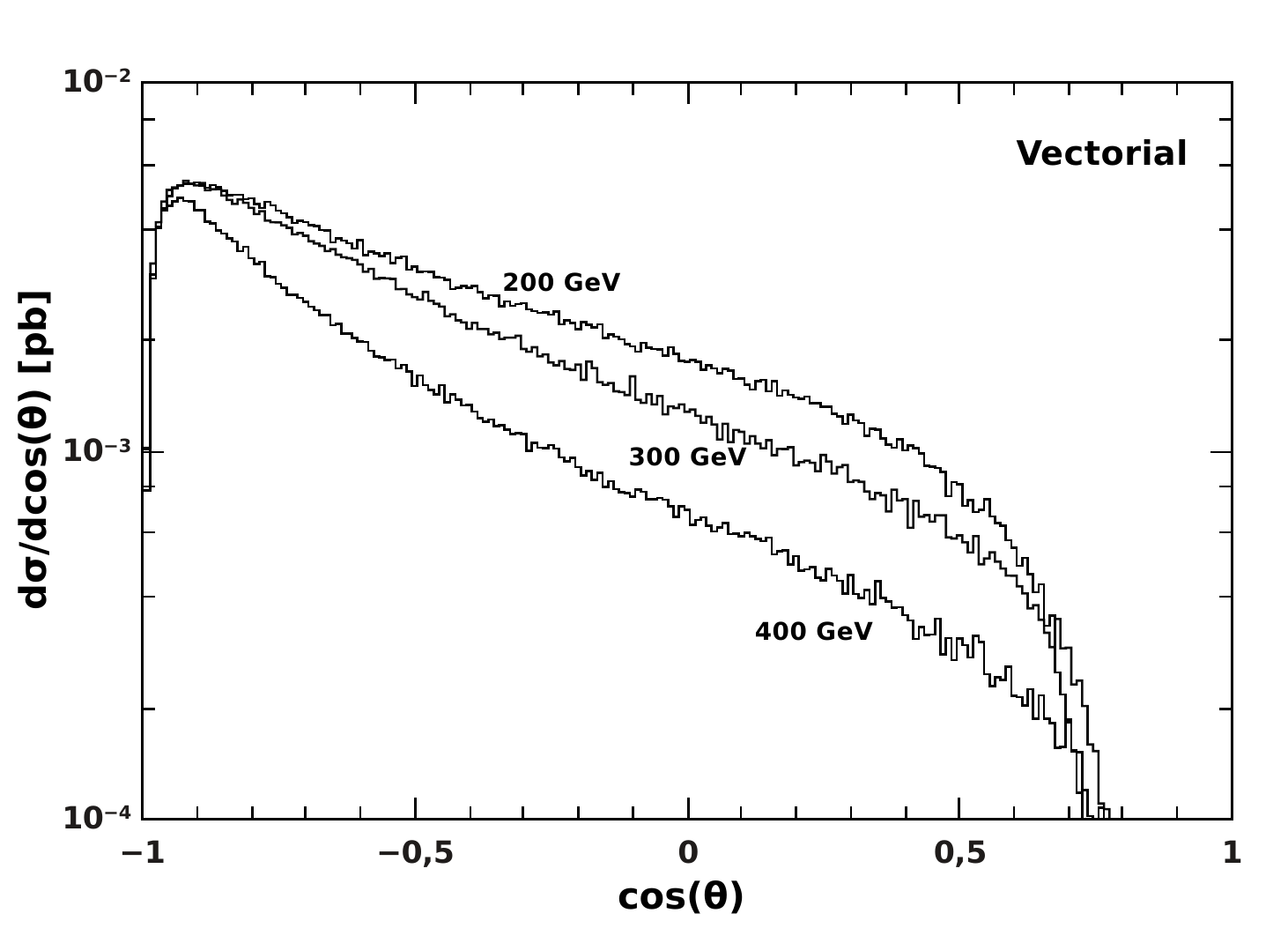}}~
\subfloat[Scalar operators contribution.]{\label{fig:dsdcos_sca}\includegraphics[totalheight=6.5cm]{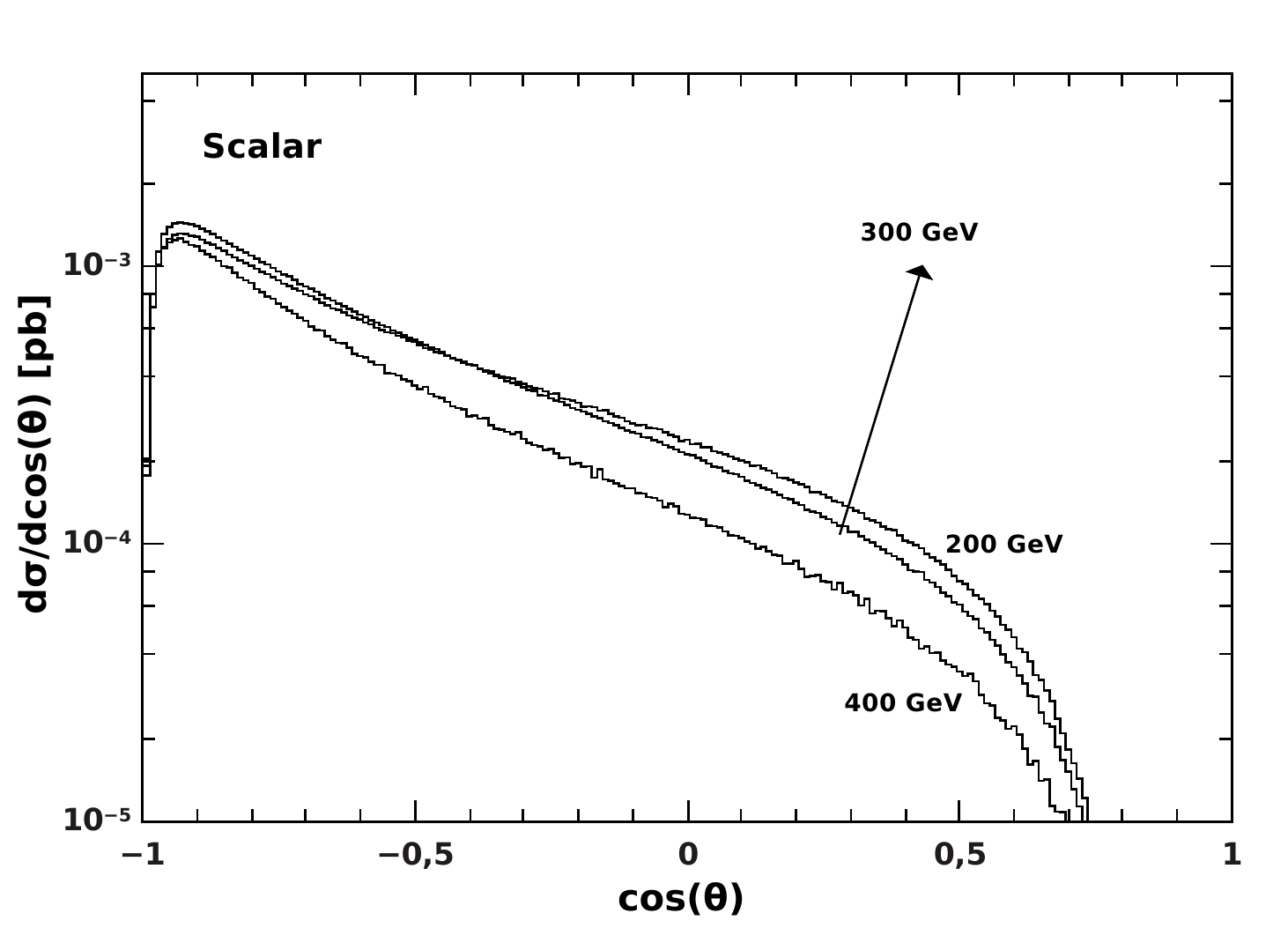}}
\caption{Angular distribution for the contribution to the cross section of vectorial and scalar operators for different Majorana neutrino masses. }
\end{figure*}

To measure the effects from the scalar operators we set the effective couplings corresponding to the vectorial operators $\alpha^{(i)}_W$ and $\alpha^{(i,j)}_{V_0}$ equal to zero, and set the value of the scalar couplings $\alpha^{(i,j)}_{S_{1,2,3}}$ in \eqref{eq:sigLam} to the values in Tab.\ref{tab:alpha-sets} corresponding to each fermion family $i,j=1,2$ considered. Similarly, to study the contribution from the vectorial operators we set the couplings $\alpha^{(i,j)}_{S_{1,2,3}}$ equal to zero, and take $\alpha^{(i)}_W=\alpha^{(i,j)}_{V_0}$ equal to the values in Tab.\ref{tab:alpha-sets}.

In Figs.\ref{fig:dsdcos_vec} and \ref{fig:dsdcos_sca} we show the angular distribution for masses $m_N= 200, ~300$ and $400$ $GeV$ for the vectorial and scalar operators respectively. The cuts presented in Fig.\ref{fig:sigbck} are applied, and the beam energies considered are $E_p=7 ~TeV$ and $E_e=150 ~GeV$ through the following. We find an asymmetric distribution for both coupling sets, favoring the backward direction, as the outgoing anti-lepton is boosted in the proton beam direction. One can also see the scalar operators contribution to the cross section is about an order of magnitude less than the vectorial contribution.

\begin{figure*}
\centering
{\includegraphics[width=0.8\textwidth]{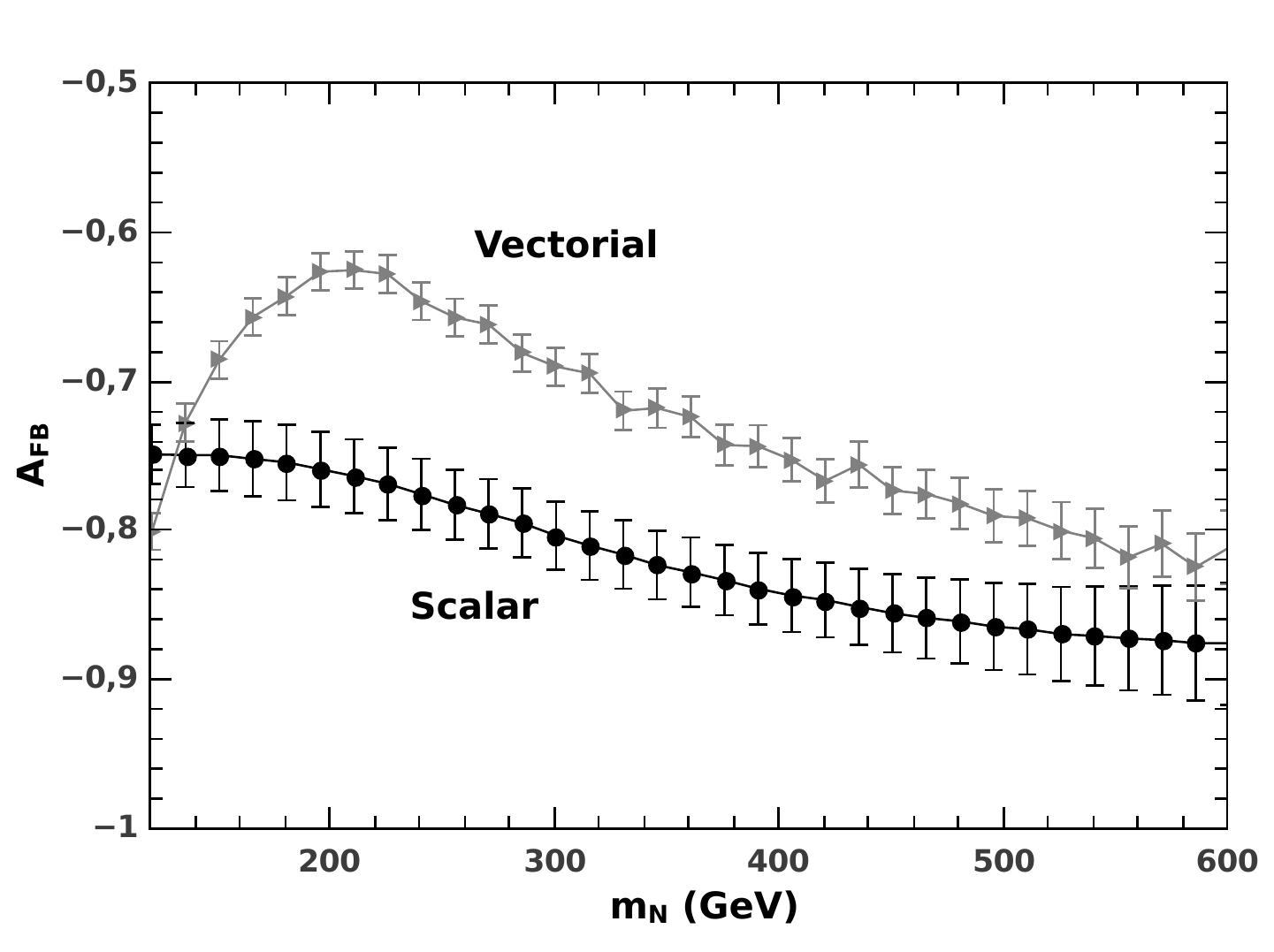}}
 \caption{\label{fig:asi} Asymmetry in the angular distribution of the final anti-lepton, with the errors estimates as defined in the text. }
\end{figure*}

In order to estimate the chances of disentangling the contributions from both operator sets, we study the angular asymmetry $A^{l^+}_{FB}$, taking into account the error 
\begin{eqnarray}\label{AsyFB}
\Delta A^{l^+}_{FB}=\sqrt{\left(\frac{\partial A^{l^+}_{FB}}{\partial N_+}\right)^2 \left( \Delta N_+ \right)^2
\, + \, \left(\frac{\partial A^{l^+}_{FB}}{\partial N_-}\right)^2 \left(\Delta N_- \right)^2}.
\end{eqnarray}
Assuming the number of events to be Poisson distributed, we write
\begin{equation}
\Delta N_+=\sqrt{N_+} \;\; \mbox{and} \;\; \Delta N_-=\sqrt{N_-}
\end{equation}
and a straightforward calculation leads to
\begin{eqnarray}
\Delta A^{l^+}_{FB} = \sqrt{ \frac{1-(A^{l^+}_{FB})^2}{N_++N_-}}.
\end{eqnarray}
The results for the $A^{l^+}_{FB}$ observable for the vectorial and scalar operators are shown in Fig.\ref{fig:asi}. Here we assume a baseline integrated luminosity of $\mathcal{L}=1 ~ab^{-1}$. We can see a clear separation between the contributions of both sets, which could help distinguishing between different kinds of new physics driving the sterile neutrino interactions.

\begin{figure*}
\centering
\includegraphics[width=0.9\textwidth]{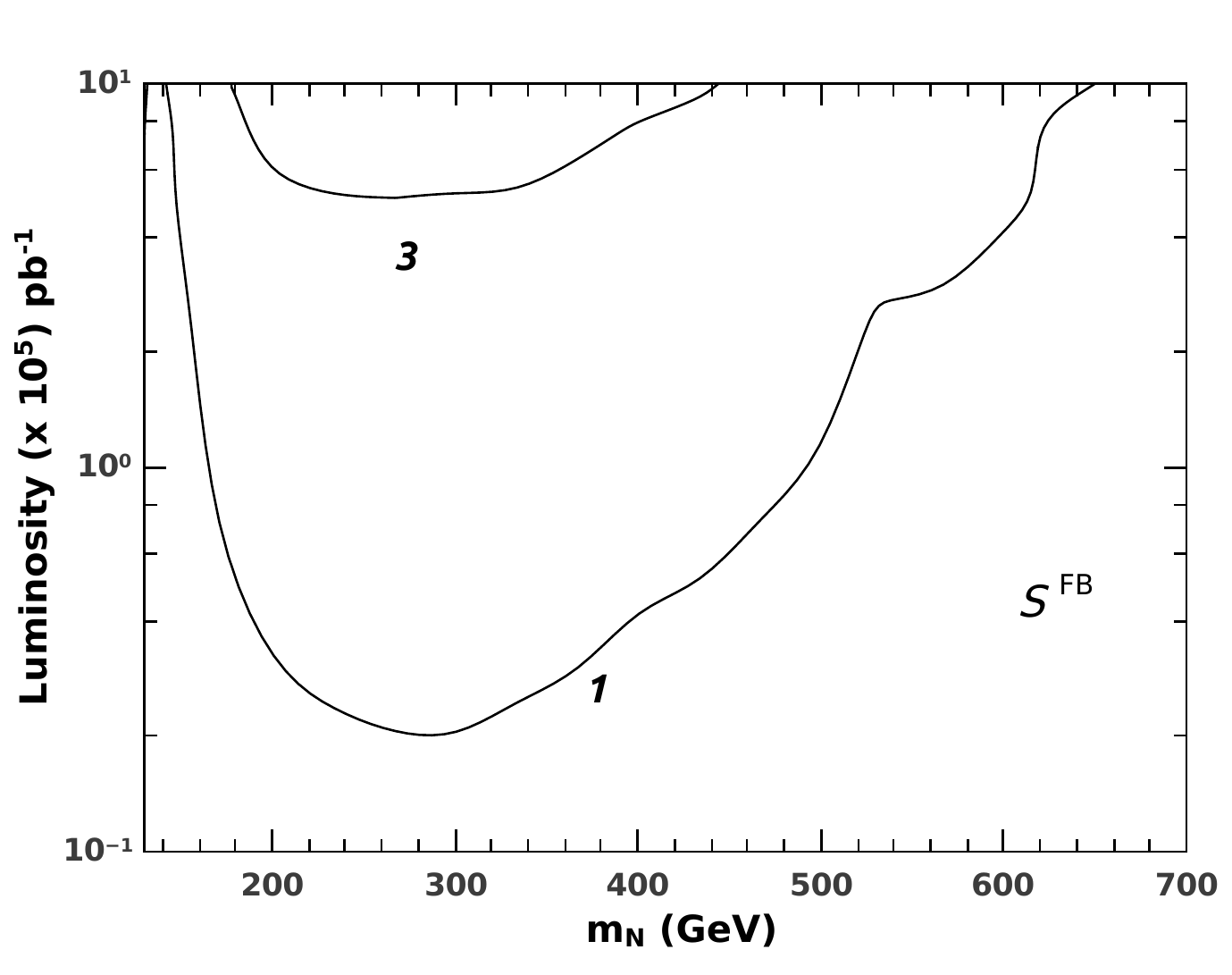}
 \caption{\label{fig:asireg1} Contour plot for the observable $\mathcal{S}^{FB}$ as defined in the text.}
\end{figure*}

We can test the ability of the forward-backward asymmetry to separate the effects of vectorial and 
scalar operators defining the quantity $\mathcal{S}^{FB}$:
\begin{equation}
\mathcal{S}^{FB}=\frac{A_{FB,vec}^{l^+}-A_{FB,sca}^{l^+}}{\Delta A_{FB,vec}^{l^+}+\Delta A_{FB,sca}^{l^+}}.
\end{equation}
In Fig.\ref{fig:asireg1} we show the contour plot for different values of $\mathcal{S}^{FB}$ in the mass-luminosity ($m_N$,$\mathcal{L}$) plane. 
The magnitude of $\mathcal{S}^{FB}$ represents the number of standard deviations between the contributions to the asymmetry from the vectorial and scalar operators. We find that the LHeC could well disentangle both effects for Majorana neutrinos in a mass range $m_N\sim 200-400~GeV$ within an operation year, when the luminosity reaches $1 ~ab^{-1}$.  

\section{\bf Initial electron polarization}\label{sec:pol}

The initial electron polarization $P_e$ can also be used to distinguish the vectorial and scalar operators contribution. It has been exploited recently in the case of the Left-Right symmetric model, where a right-polarized initial electron enhances the right handed charged current interaction \cite{Mondal:2015zba, Lindner:2016lxq}. 

\begin{figure*}
\centering
\includegraphics[width=0.8\textwidth]{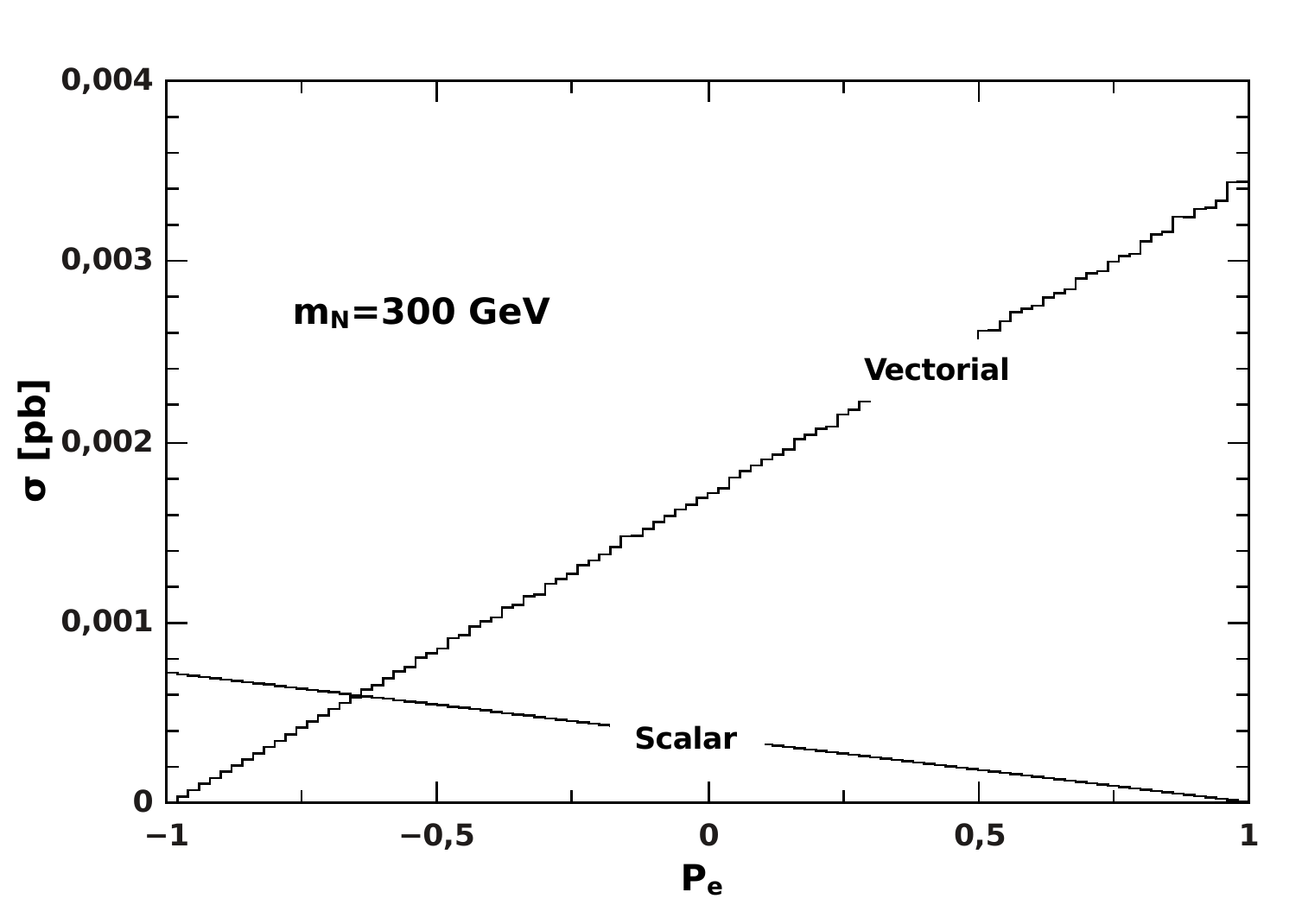}
 \caption{\label{fig:polmn300} Cross section dependence with $P_e$ for the vectorial and scalar operators. }
\end{figure*}

The cross section for the process $e^{-}p \rightarrow l_j^{+} +  3 jets$ in \eqref{eq:sigma} can be written in terms of $P_e$ as
\begin{equation}
\hat \sigma_i = \frac12(1+P_e) \hat \sigma_i^{R} + \frac12(1-P_e) \hat \sigma_i^{L}
\end{equation}
where
\begin{eqnarray}
\hat \sigma_i^{R(L)}(x s)= \int (2\pi)^4 
\delta^{(4)}(p_e+p_u-\sum_{j=1,4} 
k_j)\overline{|M_{(i)}^{R(L)}|^2} \prod_{j=1,4}\frac{d^4k_j}{2\pi^3} ~.
\end{eqnarray}
Thus, we can write again in the narrow width approximation:
\begin{eqnarray}
\overline{\vert M_{(i)}^{R(L)} \vert}^2=\left(\frac{\pi}{4m_N ~\Gamma_N ~\hat{s}}\right)\delta(k_N^2-m_N^2) \vert \Lambda_{(I),i}^{R(L)} \vert^2 
(\vert \Lambda_{(II)}^{(+)} \vert^2 + \vert \Lambda_{II}^{(-)} \vert^2)
\end{eqnarray}
where 
\begin{equation*}
\vert \Lambda_{(I),i}^R \vert^2 = \frac{16}{\Lambda^2}[(\alpha_{W}^2 \vert \Pi^{(2)}_W \vert^2)(k_d \cdot p_e)(k_N \cdot p_u)+ 
(\alpha^ 2_{V_{0}}) (k_d \cdot k_N)(p_e \cdot 
p_u) ]
\end{equation*}
and
\begin{eqnarray}
\vert \Lambda_{(I),i}^L \vert^2 = && \frac{4}{\Lambda^2}\left[ 
4(\alpha_{S_{2}}(\alpha_{S_{2}}-\alpha_{S_{3}})+\alpha^2_{S_{1}}) (k_d \cdot 
p_u)(k_N \cdot p_e)+          
\right. \nonumber \\  &&\left. \alpha_{S_{3}}(\alpha_{S_{3}}-\alpha_{S_{2}})(k_d \cdot p_e)(k_N \cdot p_u)+ 
\alpha_{S_{3}} \alpha_{S_{2}} (k_d \cdot k_N)(p_e \cdot 
p_u) \right].
\end{eqnarray}

From these expressions we can clearly see that the vectorial operators contribute to the right ($R$) part, and the scalars to the left ($L$) part of the cross section.

\begin{figure*}
\centering
\subfloat[Vectorial operators contribution.]{\label{fig:neve_vec}\includegraphics[totalheight=6.5cm]{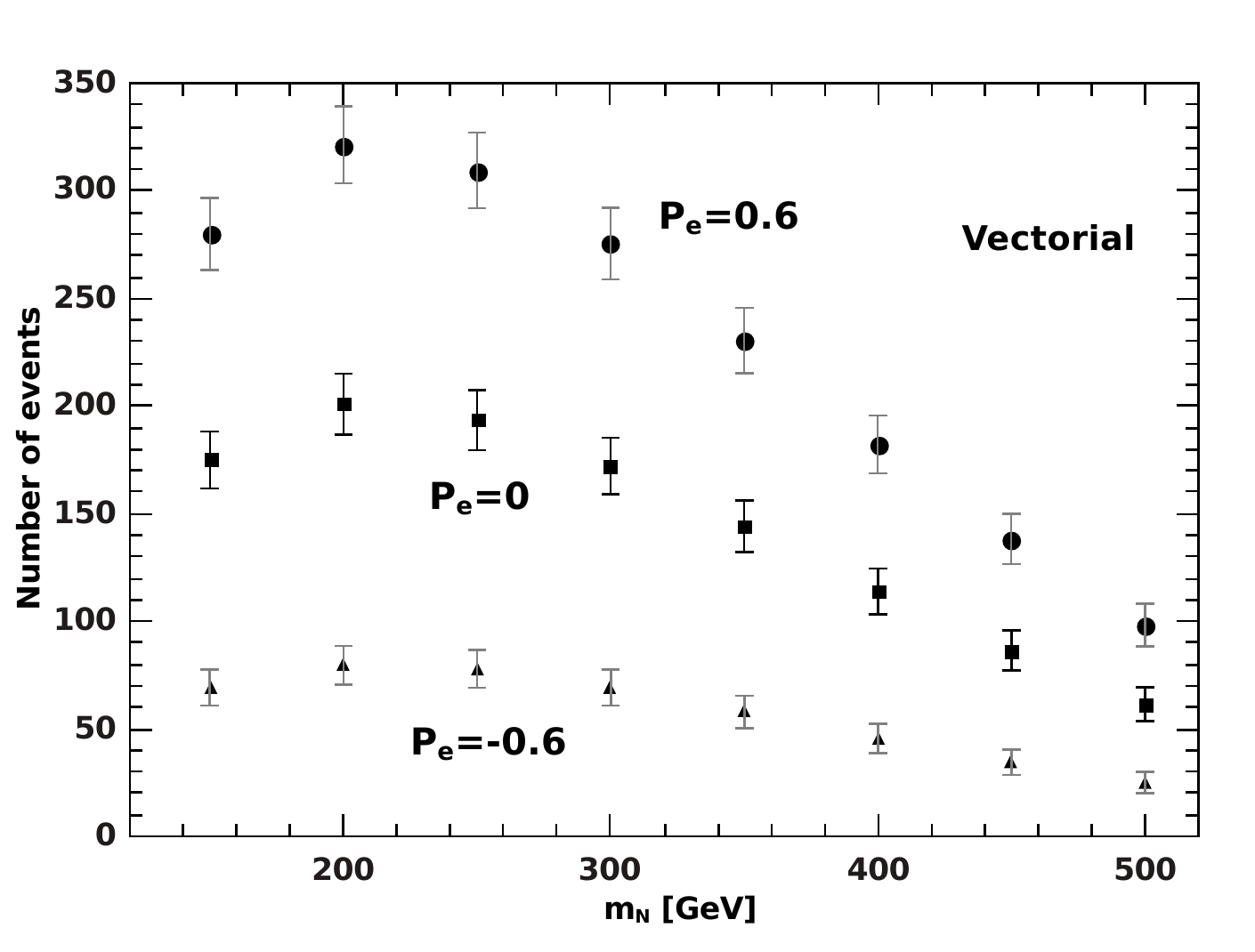}}~
\subfloat[Scalar operators contribution.]{\label{fig:neve_sca}\includegraphics[totalheight=6.5cm]{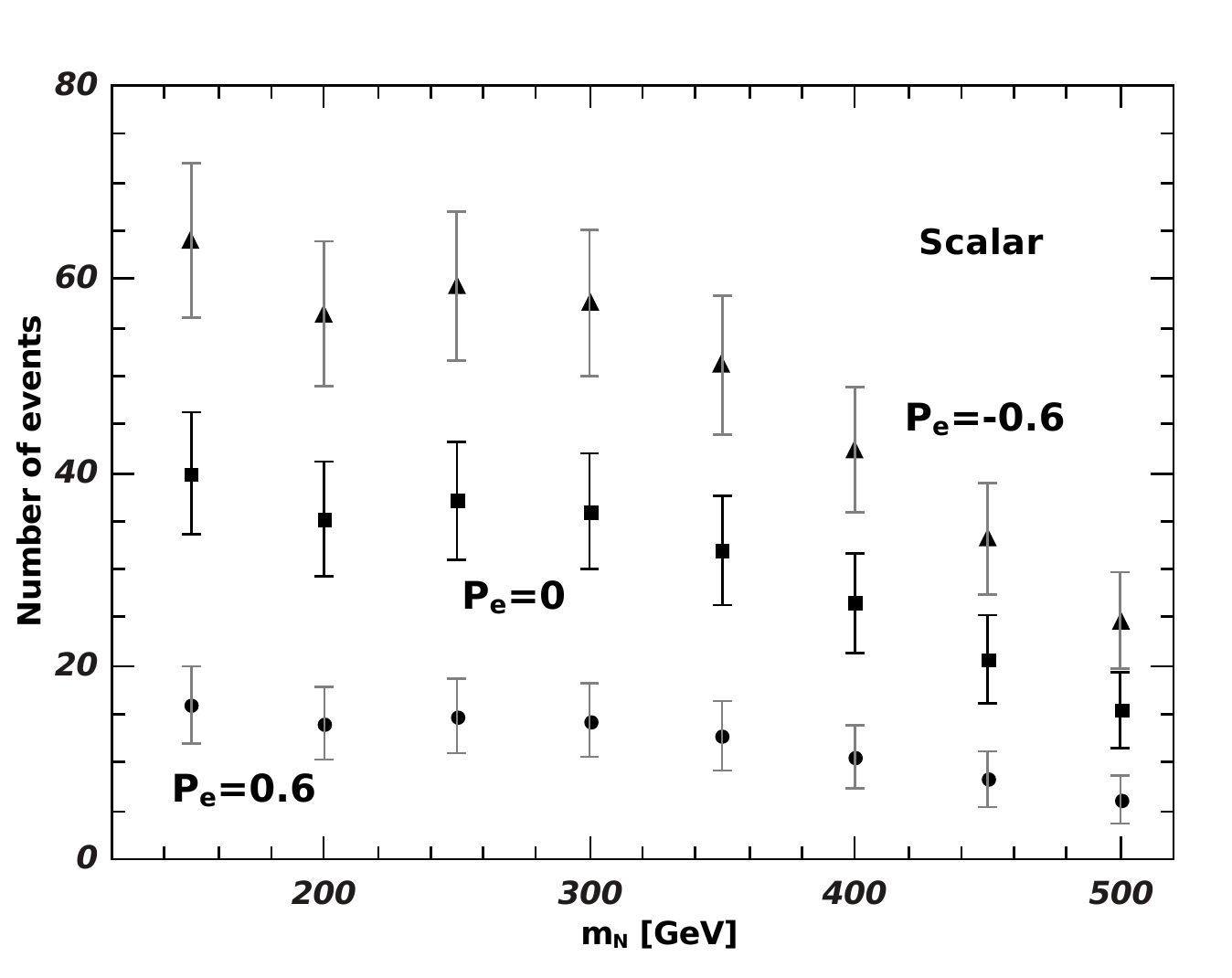}}
\caption{Number of events and error bars for the signal with $P_e=-0.6, 0$ and $~0.6 $. Here $\mathcal{L}=100 ~fb^{-1}$. }
\end{figure*}

In Fig.\ref{fig:polmn300} we show the behavior of the cross section with the polarization for $m_N=300~ GeV$ for the contribution of vectorial and scalar operators. In Figs.\ref{fig:neve_vec} and \ref{fig:neve_sca} we show the vectorial and scalar contributions respectively as a function of the Majorana neutrino mass for unpolarized ($P_e=0$), left and right-polarized electrons ($P_e=\pm 0.6$), for an integrated luminosity $\mathcal{L}=100 ~fb^{-1}$. 

\begin{figure*}
\centering
\includegraphics[width=0.9\textwidth]{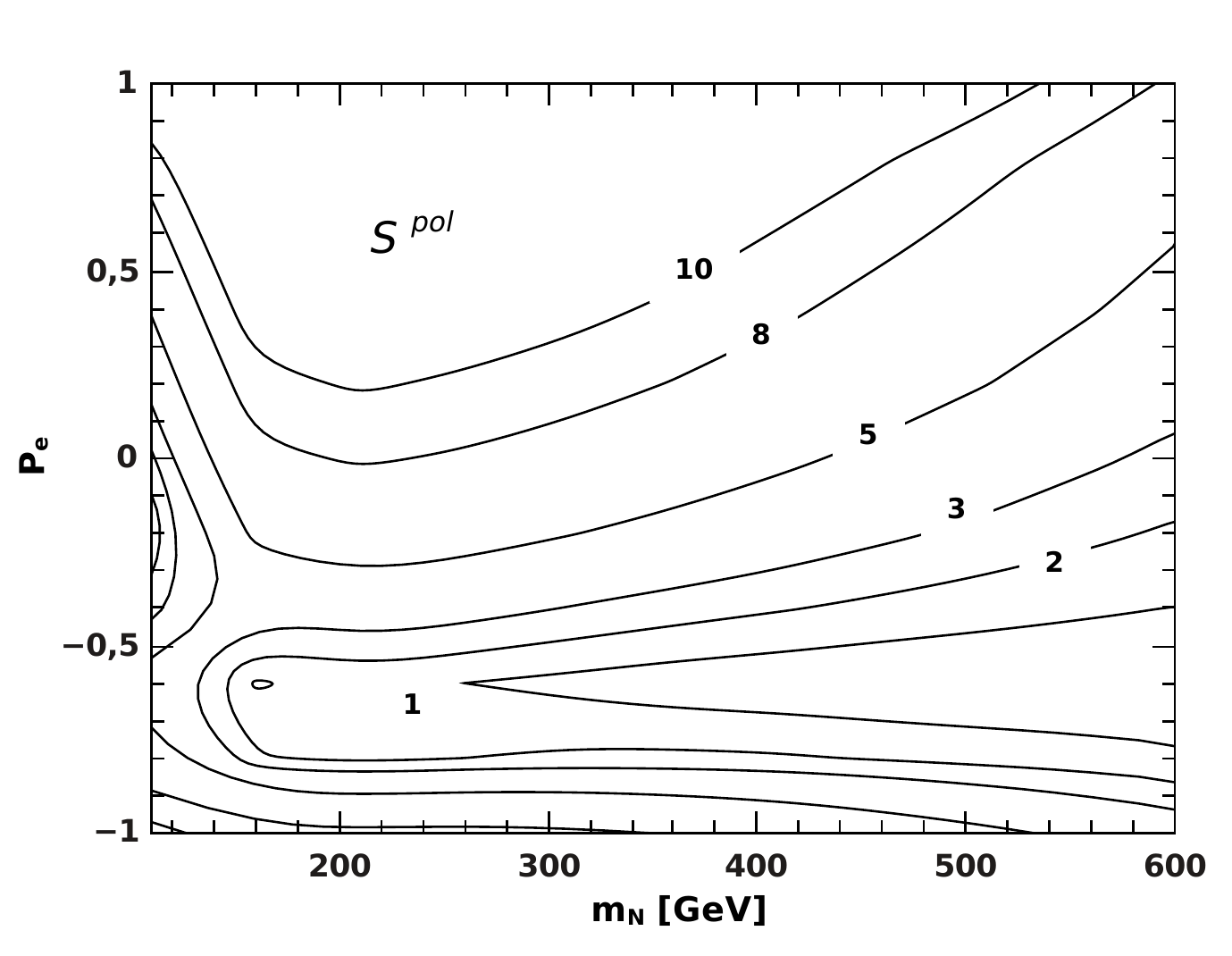}
 \caption{\label{fig:mnpolregi} Contour plot for different values of the function $\mathcal{S}^{pol}$, for $\mathcal{L}=100fb^{-1}$. }
\end{figure*}

Finally, in analogy with the last section we define the function $\mathcal{S}^{pol}$
\begin{equation}
\mathcal{S}^{pol}=\frac{N^{vec}-N^{sca}}{\sqrt{N^{vec}}+\sqrt{N^{sca}}}
\end{equation}
which represents the number of standard deviations between the numbers of events produced by the vectorial and the scalar operators contributions.
We have considered the contour plot for new function $\mathcal{S}^{pol}$ in Fig.\ref{fig:mnpolregi}, again for $\mathcal{L}=100 ~fb^{-1}$. Also here it is possible to see regions where the contributions of the different operators are considerably separated. Indeed, the polarization analysis is more promising than the angular asymmetry to distinguish the different operators contributions for a right-polarized electron beam. 

\section{Summary and conclusions}\label{sec:Concl}

The heavy neutrino effective field theory parameterizes high-scale weakly coupled physics beyond the minimal seesaw mechanism in a model independent framework, allowing for sizable lepton number violating (LNV) effects in colliders. While models like the minimal seesaw mechanism lead to the decoupling of the heavy Majorana neutrinos, predicting unobservable LNV, the effective Lagrangian framework considered in this work could serve as a means to discern between the different possible kinds of effective interactions contributing to the $e^- p \rightarrow l_j^{+} + 3 jets $ signal at the LHeC. In particular, we studied the capability of an angular observable and the initial electron polarization to disentangle the contributions of vectorial and scalar dimension 6 effective operators. 

In this paper we have calculated the total unpolarized cross section $\sigma(e^- p \rightarrow l_j^{+} + 3 jets)$ in the LHeC for different values of $m_N$ for proton and electron beams of $E_p=7 ~TeV$ and $E_e=150~ GeV$ respectively, updating the numerical values of the effective couplings $\alpha_{\mathcal{J}}$ to the existing experimental bounds, and implemented cuts in the phase space that can help to enhance the signal to background relation.

In order to discern the contribution of the different operators we calculate a forward-backward angular asymmetry for the final anti-lepton. We present our results for the asymmetry as a function of the heavy neutrino mass, for the vectorial and scalar effective interactions, considering the contour plot for the observable $\mathcal{S}^{FB}$, which measures the distance in standard deviations between the contributions to the asymmetry from the vectorial and scalar operators as a function of the heavy neutrino mass and the integrated luminosity.

We calculate the effect on the signal number of events for the vectorial and scalar interactions when the initial electron is polarized. We present the results varying the Majorana neutrino mass and the electron polarization, taking the number of events to be Poisson distributed and considering the corresponding errors.
Finally, we have studied the contour plot for the new function $\mathcal{S}^{pol}$ defined in analogy with $\mathcal{S}^{FB}$. Also in the polarization analysis it is possible to see regions where the contributions of the different operators are considerably separated.

Our findings show that electron-proton colliders, being complementary facilities to both $e^+ e^-$ and $pp$ machines which provide higher center of mass energies than the former and a cleaner environment than the latter, allow for detailed angular and polarization studies and could improve our knowledge on possible Majorana neutrinos and their interactions, which is a fundamental unsolved puzzle in particle physics.

\

{\bf Acknowledgements}

We thank CONICET (Argentina) and Universidad Nacional de Mar del
Plata (Argentina); and PEDECIBA and CSIC-UdelaR (Uruguay) for their 
financial supports.

\bibliographystyle{bibstyle.bst}
\bibliography{Bib_N_4_2018}

\end{document}